\documentclass{article}[11pt] %\twocolumn
\usepackage[utf8]{inputenc}
\usepackage{graphicx}
\usepackage{comment}
\usepackage{caption}
\usepackage{enumitem}
\usepackage{xcolor}
\usepackage{textcomp}
\usepackage{nicefrac}
\usepackage{amsfonts}
\usepackage{booktabs, makecell, tabularx}
\usepackage[a4paper, margin=0.75in]{geometry}
\usepackage{multirow}
\usepackage[font=small]{caption}
\usepackage{float}
\usepackage[normalem]{ulem}
\usepackage{amsmath}

\usepackage{ifthen}  % Required for conditional statements
\usepackage{censor}  % Required for censoring text

\newboolean{isJournalVersion}
\setboolean{isJournalVersion}{false} % Change to true for journal version

\newcommand{\conditionalCensor}[1]{%
    \ifthenelse{\boolean{isJournalVersion}}%
    {\censor{#1}}%  If isJournalVersion is true, censor the text
    {#1}%         Otherwise, display the text normally
}

\DeclareCaptionLabelSeparator{pipe}{ $\mid$ }
\begin{document} \sloppy

% Guidelines: https://pubs.rsna.org/page/ai/author-instructions#original_research

% V2 Created 02/19/23

% ------------------------------------------------------------------------
% Title Page
% ------------------------------------------------------------------------

\begin{titlepage}
    \centering
    \vskip 60pt
    \LARGE Synthetic Skull CT Generation with Generative Adversarial Networks to Train Deep Learning Models for Clinical Transcranial Ultrasound\par
    \vskip 1.5em
    \large Kasra Naftchi-Ardebili$^{1,\dagger}$, Karanpartap Singh$^{2,3,\dagger}$, Reza Pourabolghasem, Pejman Ghanouni$^{3}$, Gerald R. Popelka$^{3,4}$,  Kim Butts Pauly$^{1, 2, 3}$ \par
    \vskip 2em
    \small
    \raggedright
    $^1$Department of Bioengineering, Stanford University, Stanford, CA, 94301 \par

    $^2$Department of Electrical Engineering, Stanford University, Stanford, CA, 94301     
    
    $^3$Department of Radiology, Stanford University, Stanford, CA, 94301 \par

    $^4$Department of Otolaryngology, Stanford University, Stanford, CA, 94301
    
    \vskip 1.5em

    \normalsize
    Manuscript Type: Original Research Article \vskip 0.5em
    Funding: This work was generously supported by NIH R01 Grant EB032743. \vskip 0.5em
    Word Count: 3,324 \vskip 0.5em
    Acknowledgments: We would like to thank Jeremy Irvin and Eric Luxenberg for their helpful discussions, Ningrui Li and Fanrui Fu for advice on skull CT segmentation and preprocessing, and Jeff Elias at The University of Virginia for graciously providing 28 of the 38 human skull CTs for training SkullGAN.

    \vspace*{\fill}
    Corresponding Author: \textcolor{blue}{knaftchi@stanford.edu} \par
    $\dagger$ These authors contributed equally to this work.
\end{titlepage}

% ------------------------------------------------------------------------
% Blinded Title Page
% ------------------------------------------------------------------------

\begin{titlepage}
    \centering
    \vskip 60pt
    \LARGE Synthetic Skull CT Generation with Generative Adversarial Networks to Train Deep Learning Models for Clinical Transcranial Ultrasound \par
    \vskip 1.5em
    \large Original Research Article submitted August 18, 2023, resubmitted January 31, 2024 \par
    \vskip 1.5em
    
    \raggedright
    \Large Summary \par
    \vskip 0.5em
    \normalsize \textbf{Integration of deep learning in transcranial ultrasound can significantly optimize treatment planning in clinical settings. A major roadblock has been a lack of sufficiently large data sets of human skull CTs for training and evaluation purposes. To address this problem, we have developed SkullGAN: a deep generative adversarial network that generates large numbers of synthetic skull CT segments that are visually and quantitatively very similar to real skull CT segments and therefore can be used to train deep learning algorithms in transcranial ultrasound.}
    \vskip 1.5em

    \raggedright
    \Large Key Points \par
    \vskip 0.5em
    \normalsize 
    
    \begin{itemize}
        
        \item To address limitations in accessing large numbers of real, curated, and anonymized skull CTs to train deep learning models in transcranial ultrasound, SkullGAN was trained on 2,414 real skull CT segments from 38 healthy subjects to generate highly varied synthetic skull images. 

         \item Radiological metrics such as skull density ratio, though valid for patient evaluation for transcranial ultrasound treatments, can easily be fooled if used for statistical comparison between real and synthetic skulls. 
         
        \item Synthetic CT images generated by SkullGAN were indistinguishable from real skull CTs, as verified by t-SNE and a visual Turing test taken by expert radiologists.

    \end{itemize}

    \vskip 1.5em
\end{titlepage}

\newpage
% ------------------------------------------------------------------------
% Abstract
% ------------------------------------------------------------------------

\section*{Abstract}

\begin{itemize}[leftmargin=0pt]

    \item[] \textbf{Purpose:} Deep learning offers potential for various healthcare applications, yet requires extensive datasets of curated medical images where data privacy, cost, and distribution mismatch across various acquisition centers could become major problems. To overcome these challenges, we propose a generative adversarial network (SkullGAN) to create large datasets of synthetic skull CT slices, geared towards training models for transcranial ultrasound. With wide ranging applications in treatment of essential tremor, Parkinson’s, and Alzheimer's disease, transcranial ultrasound clinical pipelines can be significantly optimized via integration of deep learning. The main roadblock is the lack of sufficient skull CT slices for the purposes of training, which SkullGAN aims to address.
        
    \item[] \textbf{Materials and Methods:} Actual CT slices of 38 healthy subjects were used for training. The generated synthetic skull images were then evaluated based on skull density ratio, mean thickness, and mean intensity. Their fidelity was further analyzed using t-distributed stochastic neighbor embedding (t-SNE), Fréchet inception distance (FID) score, and visual Turing test (VTT) taken by four staff clinical radiologists. 
        
    \item[] \textbf{Results:} SkullGAN-generated images demonstrated similar quantitative radiological features to real skulls. t-SNE failed to separate real and synthetic samples from one another, and the FID score was 49. Expert radiologists achieved a 60\% mean accuracy on the VTT.
     
    \item[] \textbf{Conclusion:} SkullGAN makes it possible for researchers to generate large numbers of synthetic skull CT segments, necessary for training neural networks for medical applications involving the human skull, such as transcranial focused ultrasound, mitigating challenges with access, privacy, capital, time, and the need for domain expertise.

\end{itemize}

% ------------------------------------------------------------------------
% Main Body
% ------------------------------------------------------------------------

\begin{figure*}[h]
    \centering
    \includegraphics[width = \textwidth]{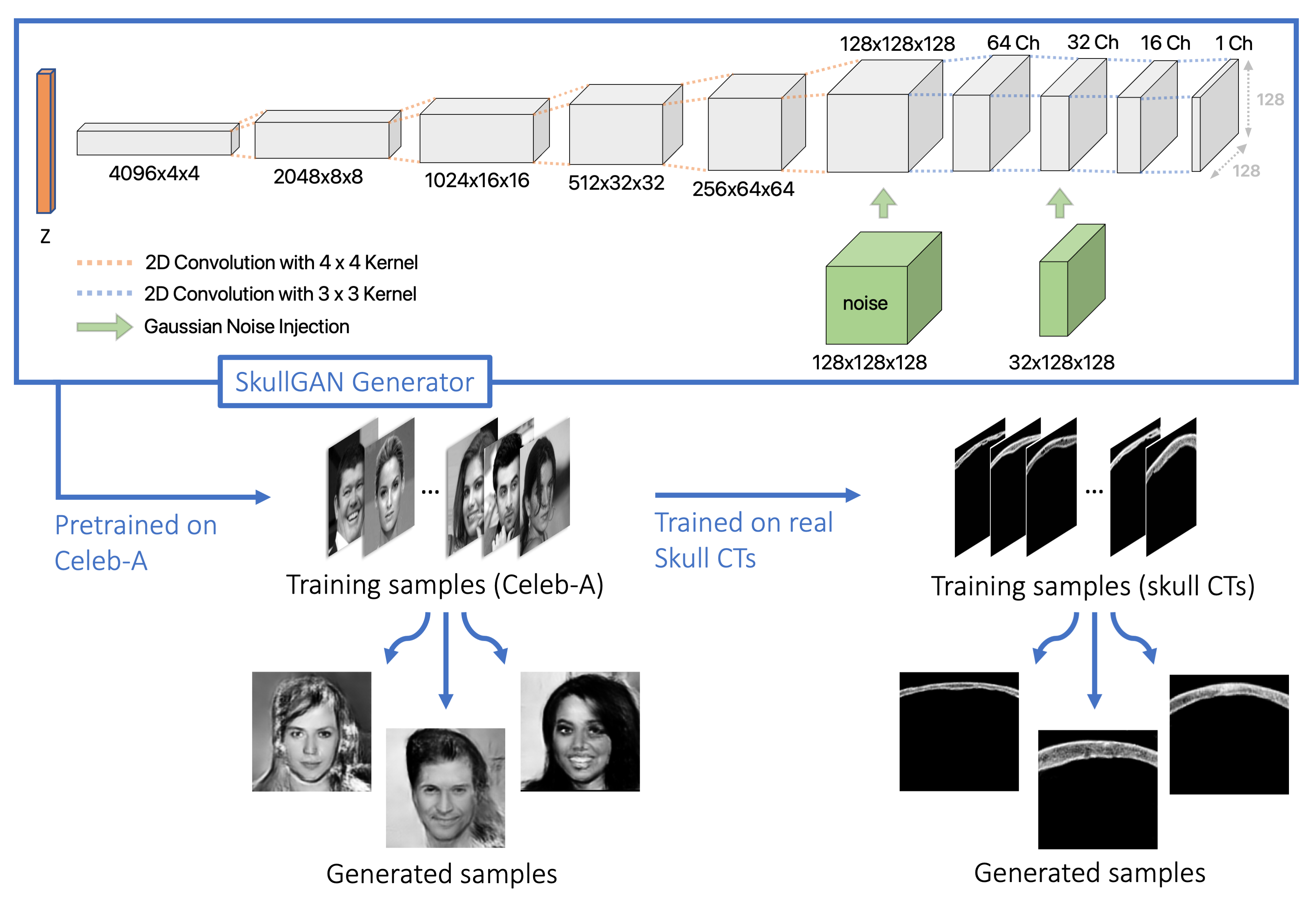}
    \caption{SkullGAN generator and training pipeline. SkullGAN was first pre-trained on the Celeb-A dataset, and then trained on human skull CTs. In contrast to random initialization of the weights for training on the human skull CTs, pre-training yielded layers with fine-tuned weights for detecting edges and resulted in better quality skull segment images, with finer definition both in contour and interior bone structure.}
    \label{architecture}
\end{figure*}

\vfill \null
\section*{Introduction}

Transcranial ultrasound stimulation (TUS), where convergent, high frequency sound waves sonicate a target deep within the brain, transcranially and noninvasively, holds the potential for treatment of a wide range of neurological disease \cite{Choi2019TherapeuticDiseases, Leinenga2016UltrasoundApplications, Piper2016FocusedReview, Elias2016ATremor, Fishman2017FocusedDisease}. Even though accounting and correcting for beam aberrations as they pass through the skull and intersect at a target inside the brain \cite{Fink1992TimePrinciples,Kyriakou2014AUltrasound.,Leung2019AUltrasound}, lends itself to being cast as an end-to-end machine learning problem, there have been little to no such attempts to date \cite{Stanziola2021AUltrasound, Shin2023Multivariable-IncorporatingSimulation}. Such a machine learning model, in contrast to the current physics-based software that suffer from an inherent trade-off between accuracy and efficiency, would be able to plan ultrasound treatments much faster while also offering higher accuracy. The main reason such an end-to-end model hasn't been attempted is a lack of a sufficiently large dataset of preprocessed real human skull CTs for the purposes of training neural networks. Anonymizing, segmentation, preprocessing, and labeling of medical images cannot be easily crowdsourced due to data privacy concerns and the need for domain expertise \cite{Litjens2017AAnalysis, Greenspan2016GuestTechnique, Roth2016ImprovingAggregation}. To address this data scarcity problem and alleviate roadblocks in integrating deep learning into TUS, we propose SkullGAN, a deep Generative Adversarial Network (GAN) that generates large numbers of synthetic skull CT segments for training deep learning models and for simulation software. We investigate whether SkullGAN-generated skull CT segments display visual similarities to real skull CTs without being exact replicas of the training set, and if they are quantitatively indistinguishable from real skull CTs.

\subsection*{Related Work}

Prior research has explored GANs for creating synthetic medical images such as cardiac MRI, liver CT, and retina images \cite{Skandarani2021GANsStudy}. In brain tumor segmentation, GANs have been used to synthesize missing T1 and FLAIR MRI sequences \cite{Conte2021GenerativeModel}, and they have been successfully deployed in generating high fidelity synthetic pelvis radiographs with orthopaedic surgeons successfully identifying synthetic images from actual with only a 55\% accuracy, very near to the expected 50\% rate \cite{Khosravi2023CreatingConcerns}. In a separate work where GAN-generated liver lesions were used to augment available real medical images for training purposes, convolutional neural network classification performance improved from $78.6\%$ to $85.5\%$ in sensitivity and from $88.4\%$ to $92.4\%$ in specificity \cite{Frid-Adar2018GAN-basedClassification}. These findings point to the potential generative models hold in generating synthetic medical images that could effectively replace real medical images for training neural networks, if enough capacity is given to the networks and adequate curated data is used to train them. However, to our knowledge, synthetic generation of human skull CT segments has never been done before.

\section*{Materials and Methods}

This study included 38 anonymized skull CT scans from healthy subjects, with 28 from the \conditionalCensor{University of Virginia}'s Department of Radiology and 10 from \conditionalCensor{Stanford}'s Department of Radiology. Informed consent was obtained from all patients, who were part of a larger prospective, multicenter study \conditionalCensor{\cite{Elias2016ATremor}}. The research was approved by the Institutional Review Board of \conditionalCensor{Stanford University} (protocol no. \conditionalCensor{IRB 32859}). 

\subsection*{Model}

SkullGAN is inspired by the Generative Adversarial Network (GAN) \cite{Goodfellow2014GenerativeNets} and the deep-convolutional GAN (DC-GAN) \cite{Radford2016DeepDCGAN}. Following the original implementation, we use binary cross entropy loss with the following loss function:

\begin{equation}
    \ell = \log{D(x)} + \log({1-D(G(z)))},
\end{equation}

\noindent where $x$ is a real skull segment sample and $z$ is a random noise vector sampled from a uniform distribution, with $D(\cdot)$ and $G(\cdot)$ denoting the discriminator and generator, respectively.

\subsection*{Network Architecture}

The generator, shown in Figure \ref{architecture}, takes a latent vector of size 200 as input and passes it through 6 sequential 2D convolutional layers. The output of the first layer has 4,096 channels, with each subsequent layer downscaling the channels by a factor of two and upscaling the features by a factor of two to yield an intermediate output of $128\times 128\times 128$. At this stage, a Gaussian noise tensor of size $128\times 128\times 128$ is added to the embedding. After reducing the channel dimension to 32 via two more convolutional layers, another Gaussian noise tensor of size $32\times 128\times 128$ is added to the embedding. This simple injection of noise at these two critical stages, determined heuristically, helps SkullGAN produce high-quality $128\times 128$ skull CT segments with well-defined trabecular pore structures.
Subsequent convolutional layers reduce the channel dimension to yield a final output of size $1\times 128\times 128$. A tanh layer constrains the final output to $[-1, 1]$, to match the normalization used for the training set. The discriminator relies on 6 convolutional layers, arranged into 6 blocks, and a sigmoid function in its final layer, to adjudicate the authenticity of its input of size $N\times 1\times 128\times 128$ ($N$ denotes the batch size). Synthetic inputs will receive scores closer to 0 whereas real inputs will receive scores closer to 1. Noise is not introduced anywhere in the discriminator network. Moreover, in contrast to the generator, instead of ReLu activations, leaky ReLu is used in the discriminator. These conditions result in 192 million parameters in the generator and 11.2 million parameters in the discriminator. Layer details are presented in Table \ref{tab:table1}.

\begin{table*}[htb!]
\resizebox{\textwidth}{!}{%
\begin{tabular}{|cccc|cccc|}
\hline
\multicolumn{4}{|c|}{Generator}                                                                                                                                                & \multicolumn{4}{c|}{Discriminator}                                                                                                                                          \\ \hline
\multicolumn{2}{|c|}{Layers}                                                                & \multicolumn{1}{c|}{Specifications}                & Output Size                 & \multicolumn{2}{c|}{Layers}                                                               & \multicolumn{1}{c|}{Specifications}                & Output Size                \\ \hline
\multicolumn{1}{|c|}{\multirow{3}{*}{Block 1}} & \multicolumn{1}{c|}{Transpose Convolution} & \multicolumn{1}{c|}{$4\times 4$ conv, stride 1, padding 0} & \multirow{3}{*}{$4096\times 4\times 4$}   & \multicolumn{1}{c|}{\multirow{2}{*}{Block 1}} & \multicolumn{1}{c|}{Convolution}          & \multicolumn{1}{c|}{$4\times 4$ conv, stride 2, padding 1} & \multirow{2}{*}{$64\times 64\times 64$}  \\ \cline{2-3} \cline{6-7}
\multicolumn{1}{|c|}{}                         & \multicolumn{1}{c|}{Batch Normalization}   & \multicolumn{1}{c|}{momentum 0.1}                  &                             & \multicolumn{1}{c|}{}                         & \multicolumn{1}{c|}{Leaky ReLU}           & \multicolumn{1}{c|}{negative slope 0.2}            &                            \\ \cline{2-3} \cline{5-8} 
\multicolumn{1}{|c|}{}                         & \multicolumn{1}{c|}{ReLU}                  & \multicolumn{1}{c|}{none}                          &                             & \multicolumn{1}{c|}{\multirow{3}{*}{Block 2}} & \multicolumn{1}{c|}{Convolution}          & \multicolumn{1}{c|}{$4\times 4$ conv, stride 2, padding 1} & \multirow{3}{*}{$128\times 32\times 32$} \\ \cline{1-4} \cline{6-7}
\multicolumn{1}{|c|}{\multirow{3}{*}{Block 2}} & \multicolumn{1}{c|}{Transpose Convolution} & \multicolumn{1}{c|}{$4\times 4$ conv, stride 2, padding 1} & \multirow{3}{*}{$2048\times 8\times 8$}   & \multicolumn{1}{c|}{}                         & \multicolumn{1}{c|}{Batch Normalization}  & \multicolumn{1}{c|}{momentum 0.1}                  &                            \\ \cline{2-3} \cline{6-7}
\multicolumn{1}{|c|}{}                         & \multicolumn{1}{c|}{Batch Normalization}   & \multicolumn{1}{c|}{momentum 0.1}                  &                             & \multicolumn{1}{c|}{}                         & \multicolumn{1}{c|}{Leaky ReLU}           & \multicolumn{1}{c|}{negative slope 0.2}            &                            \\ \cline{2-3} \cline{5-8} 
\multicolumn{1}{|c|}{}                         & \multicolumn{1}{c|}{ReLU}                  & \multicolumn{1}{c|}{none}                          &                             & \multicolumn{1}{c|}{\multirow{3}{*}{Block 3}} & \multicolumn{1}{c|}{Convolution}          & \multicolumn{1}{c|}{$4\times 4$ conv, stride 2, padding 1} & \multirow{3}{*}{$256\times 16\times 16$} \\ \cline{1-4} \cline{6-7}
\multicolumn{1}{|c|}{\multirow{3}{*}{Block 3}} & \multicolumn{1}{c|}{Transpose Convolution} & \multicolumn{1}{c|}{$4\times 4$ conv, stride 2, padding 1} & \multirow{3}{*}{$1024\times 16\times 16$} & \multicolumn{1}{c|}{}                         & \multicolumn{1}{c|}{Batch Normalization}  & \multicolumn{1}{c|}{momentum 0.1}                  &                            \\ \cline{2-3} \cline{6-7}
\multicolumn{1}{|c|}{}                         & \multicolumn{1}{c|}{Batch Normalization}   & \multicolumn{1}{c|}{momentum 0.1}                  &                             & \multicolumn{1}{c|}{}                         & \multicolumn{1}{c|}{Leaky ReLU}           & \multicolumn{1}{c|}{negative slope 0.2}            &                            \\ \cline{2-3} \cline{5-8} 
\multicolumn{1}{|c|}{}                         & \multicolumn{1}{c|}{ReLU}                  & \multicolumn{1}{c|}{none}                          &                             & \multicolumn{1}{c|}{\multirow{3}{*}{Block 4}} & \multicolumn{1}{c|}{Convolution}          & \multicolumn{1}{c|}{$4\times 4$ conv, stride 2, padding 1} & \multirow{3}{*}{$512\times 8\times 8$}   \\ \cline{1-4} \cline{6-7}
\multicolumn{1}{|c|}{\multirow{3}{*}{Block 4}} & \multicolumn{1}{c|}{Transpose Convolution} & \multicolumn{1}{c|}{$4\times 4$ conv, stride 2, padding 1} & \multirow{3}{*}{$512\times 32\times 32$}  & \multicolumn{1}{c|}{}                         & \multicolumn{1}{c|}{Batch Normalization}  & \multicolumn{1}{c|}{momentum 0.1}                  &                            \\ \cline{2-3} \cline{6-7}
\multicolumn{1}{|c|}{}                         & \multicolumn{1}{c|}{Batch Normalization}   & \multicolumn{1}{c|}{momentum 0.1}                  &                             & \multicolumn{1}{c|}{}                         & \multicolumn{1}{c|}{Leaky ReLU}           & \multicolumn{1}{c|}{negative slope 0.2}            &                            \\ \cline{2-3} \cline{5-8} 
\multicolumn{1}{|c|}{}                         & \multicolumn{1}{c|}{ReLU}                  & \multicolumn{1}{c|}{none}                          &                             & \multicolumn{1}{c|}{\multirow{3}{*}{Block 5}} & \multicolumn{1}{c|}{Convolution}          & \multicolumn{1}{c|}{$4\times 4$ conv, stride 2, padding 1} & \multirow{3}{*}{$1024\times 4\times 4$}  \\ \cline{1-4} \cline{6-7}
\multicolumn{1}{|c|}{\multirow{3}{*}{Block 5}} & \multicolumn{1}{c|}{Transpose Convolution} & \multicolumn{1}{c|}{$4\times 4$ conv, stride 2, padding 1} & \multirow{3}{*}{$256\times 64\times 64$}  & \multicolumn{1}{c|}{}                         & \multicolumn{1}{c|}{Batch Normalization}  & \multicolumn{1}{c|}{momentum 0.1}                  &                            \\ \cline{2-3} \cline{6-7}
\multicolumn{1}{|c|}{}                         & \multicolumn{1}{c|}{Batch Normalization}   & \multicolumn{1}{c|}{momentum 0.1}                  &                             & \multicolumn{1}{c|}{}                         & \multicolumn{1}{c|}{Leaky ReLU}           & \multicolumn{1}{c|}{negative slope 0.2}            &                            \\ \cline{2-3} \cline{5-8} 
\multicolumn{1}{|c|}{}                         & \multicolumn{1}{c|}{ReLU}                  & \multicolumn{1}{c|}{none}                          &                             & \multicolumn{1}{c|}{Block 6}                  & \multicolumn{1}{c|}{Convolution}          & \multicolumn{1}{c|}{$4\times 4$ conv, stride 1, padding 0} & $1\times 1$                        \\ \hline
\multicolumn{1}{|c|}{Block 6}                  & \multicolumn{1}{c|}{Transpose Convolution} & \multicolumn{1}{c|}{$4\times 4$ conv, stride 2, padding 1} & $128\times 128\times 128$                 & \multicolumn{4}{c|}{\multirow{5}{*}{}}                                                                                                                                      \\ \cline{1-4}
\multicolumn{1}{|c|}{\multirow{4}{*}{Block 7}} & \multicolumn{1}{c|}{Convolution}           & \multicolumn{1}{c|}{$3\times 3$ conv, stride 1, padding 1} & $64\times 128\times 128$                  & \multicolumn{4}{c|}{}                                                                                                                                                       \\ \cline{2-4}
\multicolumn{1}{|c|}{}                         & \multicolumn{1}{c|}{Convolution}           & \multicolumn{1}{c|}{$3\times 3$ conv, stride 1, padding 1} & $32\times 128\times 128$                  & \multicolumn{4}{c|}{}                                                                                                                                                       \\ \cline{2-4}
\multicolumn{1}{|c|}{}                         & \multicolumn{1}{c|}{Convolution}           & \multicolumn{1}{c|}{$3\times 3$ conv, stride 1, padding 1} & $16\times 128\times 128$                  & \multicolumn{4}{c|}{}                                                                                                                                                       \\ \cline{2-4}
\multicolumn{1}{|c|}{}                         & \multicolumn{1}{c|}{Convolution}           & \multicolumn{1}{c|}{$3\times 3$ conv, stride 1, padding 1} & $1\times 128\times 128$                   & \multicolumn{4}{c|}{}                                                                                                                                                       \\ \hline
\multicolumn{1}{|c|}{Final Layer}              & \multicolumn{1}{c|}{Normalizing Layer}     & \multicolumn{1}{c|}{Hyperbolic Tangent}            & $1\times 128\times 128$                   & \multicolumn{1}{c|}{Final Layer}              & \multicolumn{1}{c|}{Classification Layer} & \multicolumn{1}{c|}{Sigmoid}                       & $1\times 1$                        \\ \hline
\end{tabular}%
}
\caption{Generator and discriminator architectures in SkullGAN. In the generator, Gaussian noise is added to the outputs of Block 6 and second layer of Block 7.}
\label{tab:table1}
\end{table*}

\subsection*{CT Imaging and Segmentation}

\begin{figure}[htb!]
    %\twocolumn
    \centering
    \includegraphics[width=0.48\textwidth]{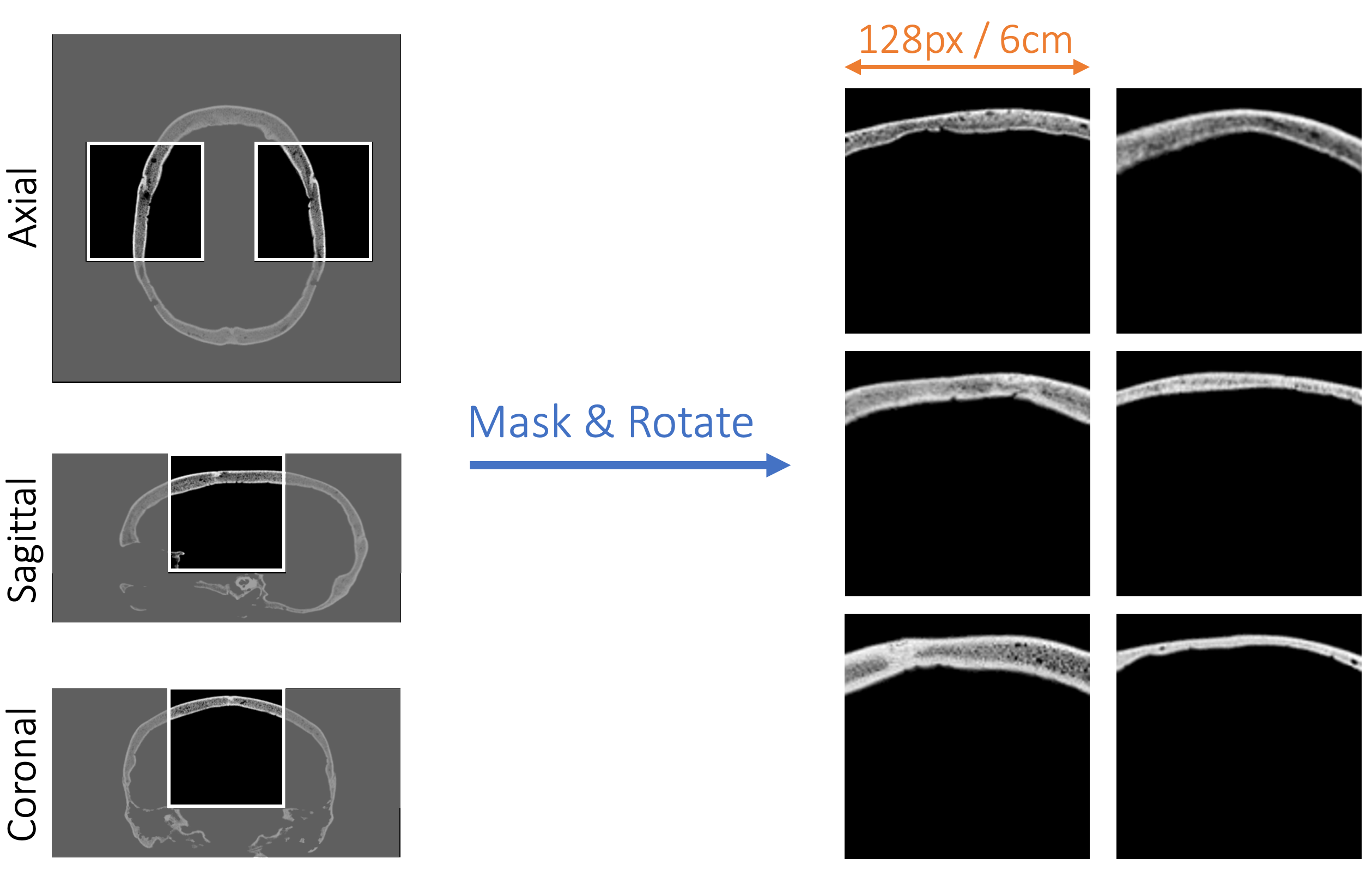}
    \caption{SkullGAN training set preparation. After segmentation, the slices were masked and rotated where necessary. To account for both the left and the right temporal bones, two segments were taken from each axial slice. This resulted in a training set of 2,414 2D horizontal skull segments.}
    \label{inputs}
\end{figure}

All 38 skull CT scans were taken at 120 keV on GE scanners with axial slicing, a 0.625 mm slice thickness, and the Bone Plus kernel. The skull CTs were segmented using ImageJ and Slicer \cite{Schneider2012NIHAnalysis, Fedorov20123DNetwork} to remove the brain and artifacts outside the skull. Slices were selected at an interval of 3.1 mm (every 5 slices) in the axial, coronal, and sagittal planes. As such, our training set consisted of slices from the temporal and parietal bones. Given the high variability in skull structure within subjects, both the left and the right temporal bones were extracted from every axial plane, without fear of biasing the dataset. The choice of temporal and parietal bones for training the model was inspired by the fact that in transcranial ultrasound settings, transducers are typically placed on these locations and not on the frontal or occipital bones. Slices were masked in MATLAB \cite{2021MATLABR2021a} to produce a final dataset of 2,414 2D skull segments for training (Figure \ref{inputs}).

\subsection*{Datasets}

\begin{figure*}[htb!]
    \centering
    \includegraphics[width = \textwidth]
    {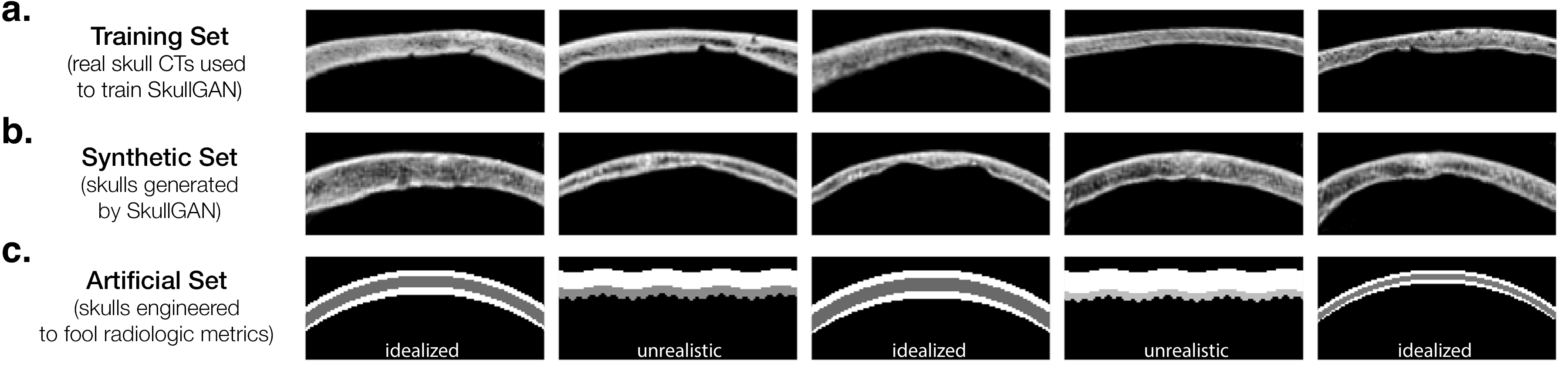}
    \caption{Five cropped examples from each dataset. \textbf{a.} Training Set: real skull CT segments. \textbf{b.} Synthetic Set: skull CT segments generated by SkullGAN. \textbf{c.} Artificial Set: idealized and unrealistic fake skull CT segments deliberately engineered to look unlike any real skull segments and yet fool quantitative radiological assessment metrics.}
    \label{examples}
\end{figure*}

The training and evaluation of SkullGAN involve multiple sets of images, each catering to unique stages in the training and assessment of the model. The first step involved pre-training SkullGAN on a large dataset of human faces to tune its weights for generating detailed structures, so that subsequent training on skull CTs does not begin with a random initialization. The second step entails the main training process, where SkullGAN is trained on real human skull images. The output of SkullGAN is referred to as the synthetic set, which is a collection of generated skull CT images. Example images from each set are shown in Figure \ref{examples}. 

\begin{itemize}[leftmargin=0in]

\item[] \textbf{pre-training set}: 100,000 cropped and rescaled $128 \times 128$ celebrity images (Celeb-A dataset) \cite{Liu2018Large-scaleDataset}. This dataset,  well-known in computer vision literature \cite{Yu2018GenerativeAttention, Feng2021WhenSize}, was used during pre-training to allow the model to learn fundamental facial structures and features, which facilitated its subsequent learning of human skulls during the main training phase.

\item[] \textbf{training set}: 2,414 real human skull CT segments used to train SkullGAN.

\item[] \textbf{synthetic set}: 1,000 2D synthetic skull segments generated by SkullGAN.

\item[] \textbf{artificial set}: 500 \emph{idealized} skull segments, engineered to represent the simplest model of the skull, and 500 \emph{unrealistic} skull segments, purposefully engineered to look ostensibly unreal. Although visually distinct from real skull images, these artificial images are designed to fool common quantitative radiological metrics, illustrating the potential limitations of these common metrics when assessing the authenticity of synthetic skulls. 
\end{itemize}

\subsection*{Training}

SkullGAN was pre-trained on 100,000 Celeb-A samples \cite{Liu2018Large-scaleDataset} for 10 epochs and then trained on the training set for 1,000 epochs. Pre-training greatly improved the resolution of the synthetic set, while reducing the number of iterations required for convergence during training (Figure \ref{pretrain_effect} in the Appendix). The training set was normalized to a range of [-1, 1]. Both the generator and discriminator networks used mini-batch training with a batch size of 64. Optimization was performed using the Adam optimizer \cite{Kingma2015Adam:Optimization} with no weight decay and $\beta=0.5$.

To stabilize the networks and improve the synthetic skulls, several techniques were applied. For the discriminator, we used label smoothing by assigning soft labels of 0.9 (instead of 1.0) for real samples to encourage stabilization. Starting with a learning rate of $3\times 10^{-3}$, a dynamic learning rate reduction on plateau \cite{PyTorch2019PyTorchDocumentation} was used for both networks: For the generator, a decay factor of 0.5 and patience of 1,000 iterations was determined through our
parameter search, while for the discriminator, a decay factor
of 0.8 and patience of 1,000 iterations were used. Each
network was allowed to update only if its loss in the last
batch was lower than a heuristically-determined value of
90\%. Gaussian-blurred real samples were gradually introduced to the discriminator as fake samples (0 labeled) past the 15$^{th}$ epoch, until a ratio of \nicefrac{1}{2} blurred real samples and \nicefrac{1}{2} fake samples was reached. This ratio was then kept constant. Introduction of blurred reals as fakes was to guide the model to emphasize high resolution skull pore structures rather than learn only the inner and outer contours. 

\subsection*{Hardware}

SkullGAN was developed and trained on 2 NVIDIA A100 40GB GPUs, running on a machine-learning optimized Google Cloud Platform \cite{GoogleCloudPlatformGCP2019GoogleGCP} instance. 

\subsection*{Validation}

\subsubsection*{Skull Density Ratio (SDR)}

A widely used quantitative metric in determining whether essential tremor patients are candidates for ablation using MR-guided focused ultrasound is SDR \cite{Tsai2021TheThalamotomy, DSouza2019ImpactTremor, Yuen2022HyperostosisThalamotomy}. SDR was calculated for each skull segment by taking 32 vertical rays down the segment, spaced approximately 1.8 mm apart, and then computing the mean ratio of the minimum to maximum pixel intensities along each ray:

\begin{equation}
\mathrm{SDR_j}= \frac{1}{32}\sum_{i=1}^{32}\frac{\min({S_{ji}})}{\max(S_{ji})},
\end{equation}

\noindent where $\mathrm{SDR_j}$ denotes the SDR for skull $j$, and $S_{ji}$ refers to the $i^{th}$ vertical ray down the $j^{th}$ skull segment. 

\subsubsection*{Mean Thickness}

The mean thickness of each skull segment was calculated by averaging the thickness through 32 rays down the image, spaced approximately 1.8 mm apart: 

\begin{equation}
\mathrm{MT_j} = \frac{1}{32}\sum_{i=1}^{32} \rho \times T_{i},
\end{equation}

\noindent where $\mathrm{T_i}$ denotes the thickness--in pixels--of the $i^{th}$ ray, and $\rho$ denotes the CT resolution in $\mathrm{\nicefrac{mm}{pixel}}$.

\subsubsection*{Mean Intensity}

The mean intensity for each skull segment was calculated by first thresholding the image to ignore Hounsfield Unit (HU) values of 10 or less (the background). The intensity of each pixel was then averaged to obtain the mean intensity of the skull segment: 

\begin{equation}
\mathrm{MI_j}= \frac{1}{N_j}\sum_{\mathrm{x, y}} I(x,y)\cdot\mathbf{1}_{I(x,y) > 10 \> \mathrm{HU}}
\end{equation}

\noindent where $\mathrm{MI_j}$ denotes the mean intensity for skull segment $j$, and $N_j$ is the total number of pixels in segment $j$ that meet the thresholding requirement of $>$ 10 HU.

\subsubsection*{t-Distributed Stochastic Neighbor Embedding}

To assess the similarities or differences between large samples of real, synthetic, and artificial skull slices, we employed t-distributed stochastic neighbor embedding (t-SNE) \cite{vanderMaaten2008VisualizingT-SNE}. The t-SNE algorithm constructs a probability distribution for pairs of objects based on their similarity, both in the original high-dimensional space and in a lower dimensional representation, and then iteratively solves for a mapping between the two distributions, thus representing the high-dimensional objects in a lower-dimensional, visually interpretable manner. We applied t-SNE to assess the separability of the three sets—real, synthetic, and artificial—each comprising 1,000 data points. The goal was to determine if distinct clusters could be observed in a 2D subspace..

\subsubsection*{Fréchet Inception Distance Score}
One common method in assessing the quality and diversity of the generated images is the Fréchet Inception Distance (FID) score \cite{Heusel2017GANsEquilibrium}. It compares the mean and standard deviation of the deepest (final) layer of the Inception-v3 model \cite{Szegedy2015RethinkingVision} applied to batches of real and generated images. The final layer of this model is a 1,000-dimensional feature vector, where each element corresponds to one of the thousand different classes in the ImageNet dataset \cite{Deng2009ImageNet:Database}. We generated 10,000 synthetic skull segments using SkullGAN, and sampled 10,000 real CT segments from our 38 human skull CTs. Pre-processing of these samples included resizing to $299\times 299\times 3$, where the channels were scaled to have means $[0.485, 0.456, 0.406]$ and standard deviations $[0.229, 0.224, 0.225]$, as per original implementation of FID score. Application of Inception-v3 model on these pre-processed real and synthetic samples resulted in $F_{r}\in \mathbb{R}^{1,000\times 10,000}$ and $F_{s}\in \mathbb{R}^{1,000\times 10,000}$ feature matrices, with class-wise mean vectors $\mu_{r}\in\mathbb{R}^{1,000}$ and $\mu_{s}\in\mathbb{R}^{1,000}$, respectively. With covariance matrices $\Sigma_{r}\in\mathbb{R}^{1,000\times 1,000}$ and $\Sigma_{s}\in\mathbb{R}^{1,000\times 1,000}$ constructed from these mean vectors, the FID score is computed as follows:

\begin{equation}
\text{FID} = (\mu_r-\mu_s)^T(\mu_r -\mu_s) + \text{Tr}\big( \Sigma_r + \Sigma_s - 2\sqrt{\Sigma_r \odot \Sigma_s} \big),
\end{equation}

\noindent with $\text{Tr}(\cdot )$ representing the trace operator and $\odot$ being the element-wise product.

\subsubsection*{Visual Turing Test}
The most robust measure of accuracy and fidelity of generative models is to have human graders label their outputs as real or synthetic. The specific visual Turing test (VTT) \cite{Geman2015VisualSystems} we designed consisted of 25 real and 25 synthetic samples, with 6 identical duplicates in each category, presented to 4 staff clinical radiologists, in random order, without the option of viewing or editing past responses. Their performance report comprised true positive rate (TPR), false positive rate (FPR), switch rate (the number of times they switched their labels of the identical duplicates), as well as the average time they spent on studying a synthetic sample versus a real sample before deciding on a label:

\begin{equation}
    TPR = \frac{TP}{FN + TP} = \frac{TP}{25}
\end{equation}

\begin{equation}
    FPR = \frac{FP}{TN + FP} = \frac{FP}{25}
\end{equation}

\begin{equation}
    \text{Accuracy} = \frac{TP + TN}{FN + TP + TN + FP} \times 100= \frac{TP + TN}{50}\times 100
\end{equation}

A 50\% accuracy marks the theoretical optimum where the expert graders resort to guessing, and corresponds to high fidelity of the synthetic skull segments.       

\subsection*{Memory GAN}

A common challenge in training GANs is ``Memory GAN,'' in which the network simply memorizes the training set \cite{Nagarajan2019TheoreticalGANs}. To test for this failure mode and verify the uniqueness of our SkullGAN-generated segments, we compared all of our training set to an equally sized batch of our synthetic set. To find the closest real counterparts to the synthetic segments, we searched for the minimum distance between synthetic and real samples, computed once via scale-invariant feature transform (SIFT) \cite{Lowe1999ObjectFeatures}, once with simple pixel-wise mean squared error (MSE), and lastly via cosine similarity of their ResNet50 \cite{HeDeepRecognition} feature vectors. 

% ------------------------------------------------------------------------

% ------------------------------------------------------------------------

\section*{Results}

Figure \ref{examples} displays example images from the training set, the synthetic set generated by SkullGAN, and the artificial set, separated into idealized and unrealistic varieties. The total training time for SkullGAN was approximately 2 hours. Generation time averaged 2.95 seconds for 2,500 images and 9.6 minutes for 100,000 images.

\begin{figure*}[htb!]
    \centering
    \includegraphics[width = \textwidth]{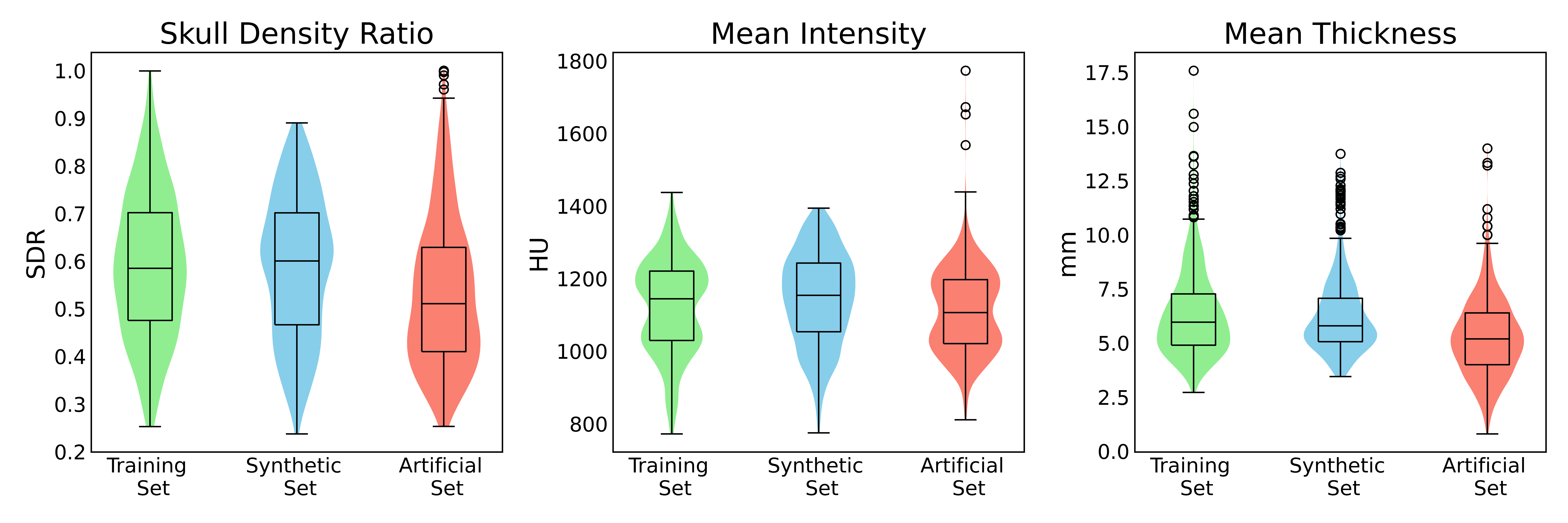}
    \caption{Violin plots of the quantitative radiological metrics for the training, synthetic, and artificial sets. What is noteworthy is that we can engineer artificial skulls that are ostensibly unrealistic and still match the training set (real skull CTs) across the three radiological metrics of skull density ratio, mean intensity, and mean thickness. In fact, we can go as far as matching the shapes of the distributions: the bimodal distribution of mean intensity for the artificial set resembles that of the training set.}
    \label{distributions}
\end{figure*}

\subsection*{Susceptibility of the Quantitative Radiological Metrics to Failure}

While quantitative radiological metrics such as SDR, mean thickness, and mean intensity are viable measures to compare real skull CTs for the purpose of clinical evaluation \cite{Elias2016ATremor}, they can fail when assessing the authenticity of synthetic skull CTs. As shown in Figure \ref{distributions}, even the artificial skull CT segments engineered to match the SDR, mean thickness, and mean intensity distributions of real skull CT segments can easily fool these metrics, even though they are visually clearly fake. Therefore, other extensive methods were employed to analyze the separability of these datasets.

\subsection*{Visual Clustering of the Data Sets via t-SNE}

\begin{figure*}[!htb]
    \centering
    \includegraphics[width = 0.8\textwidth]{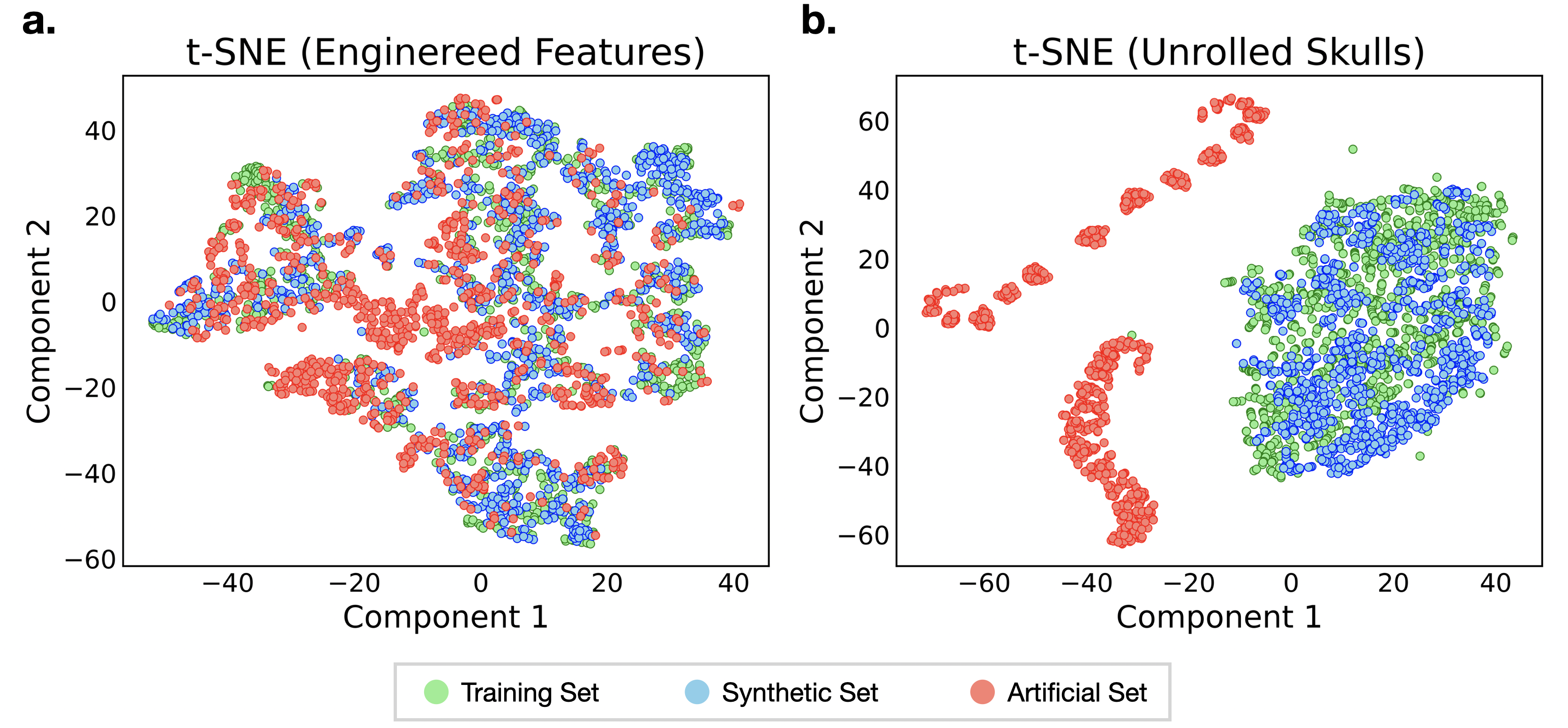}
    \caption{Separability of the datasets. \textbf{a.} Visual representation of t-SNE applied to the radiological features shown in Figure \ref{distributions}. No discernible clustering is seen, and the datasets appear inseparable with this method. \textbf{b.} Visual representation of t-SNE applied to the unrolled skulls, where each image is represented as a vector of size 16,384. The artificial set is clearly separated into clusters by t-SNE (one for the unrealistic models and another for the idealized models), while the other distributions remain inseparable.}
    \label{t-SNE}
\end{figure*}

When applying t-SNE to the distributions of radiologic metrics for all of the data points (Figure \ref{t-SNE}a), we observed no separability, confirming the inadequacy of these features in authenticating the real skull CT segments from the synthetic and artificial sets. However, once we break free from these limiting radiological metrics and instead unroll every sample into a vector of length $128\times 128 = 16,384$, t-SNE treats every entry (pixel) in this vector as a feature. Applying t-SNE to the $3,000\times 16,384$ matrix where the rows represent training, synthetic, and artificial skull sets in batches of 1,000, we observed a clear separate clustering of the artificial set. Interestingly, within the artificial set, the unrealistic segments were in a clearly separate cluster than the idealized segments (Figure \ref{t-SNE}b). Despite this, t-SNE still fell short of separating the training set from the synthetic set, further strengthening the claim that SkullGAN-generated synthetic images were quantitatively indistinguishable from the real skull CT segments. 

\subsection*{Inconclusivity of the FID Score}
Comparing a batch of 10,000 real skull CT segments and 10,000 synthetic segments resulted in an FID score of 49. Although a lower score indicates similarity between the real and synthetic images and could be used as a proxy for generated-image quality, a higher score does not necessarily mean lack of quality and could in turn indicate diversity in the synthetic set. A further complication in interpreting our FID score is that Inception-v3 is trained on the ImageNet dataset that has no skull CTs. In fact, Inception-v3 classified the entirety of the 10,000 real set as well as the entirety of the 10,000 synthetic set as the $111^{th}$ class of the ImageNet dataset, which corresponds to nematodes. 

\subsection*{Visual Turing Test}
As a definitive test of the quality and fidelity of the synthetic skulls generated by SkullGAN, we presented our quiz of 50 questions to 4 staff clinical radiologists. On average, they achieved a 60\% accuracy in correctly labeling the samples presented to them. Additionally, on average, they spent 7.16 seconds on studying a synthetic sample and 8.05 seconds in studying a real sample, before submitting their decision. This indicated that the synthetic sets were realistic enough that the expert graders actually ended up spending similar amount of time on them as they did on real samples before coming down with a decision. Additionally, by placing exact duplicates in the quiz, we were able to assess the confidence the experts had in their judgments. On average, in 25\% of the duplicate samples, they switched their labels. A summary performance report is presented in Table \ref{tab:table2} as well as Figure \ref{SkullGAN-Quiz} in the Appendix.

\begin{table*}[htb!]
\resizebox{\textwidth}{!}{
\begin{tabular}{l|l|l|l|l|l|l|l|}
\cline{2-8}
                               & Total Test Time & \begin{tabular}[c]{@{}l@{}}Average Time on\\ Synthetic Samples\end{tabular} & \begin{tabular}[c]{@{}l@{}}Average Time on\\ Real Samples\end{tabular} & Overall Accuracy & \begin{tabular}[c]{@{}l@{}}True Positive\\ Rate\end{tabular} & \begin{tabular}[c]{@{}l@{}}False Positive\\ Rate\end{tabular} & Switch Rate \\ \hline
\multicolumn{1}{|l|}{Expert 1} & 8.16 min        & 9.03 s                                                                    & 10.55 s                                                              & 58.0 \%          & 0.48                                                         & 0.36                                                          & 16.67 \%    \\ \hline
\multicolumn{1}{|l|}{Expert 2} & 8.99 min        & 9.45 s                                                                    & 12.11 s                                                              & 66.0 \%          & 0.60                                                         & 0.32                                                          & 16.67 \%    \\ \hline
\multicolumn{1}{|l|}{Expert 3} & 4.08 min        & 4.36 s                                                                    & 5.44 s                                                               & 58.0 \%          & 0.32                                                         & 0.20                                                          & 16.67 \%    \\ \hline
\multicolumn{1}{|l|}{Expert 4} & 4.13 min        & 5.82 s                                                                    & 4.10 s                                                               & 58.0 \%          & 0.48                                                         & 0.36                                                          & 50.0 \%     \\ \hline
\multicolumn{1}{|l|}{\textbf{Average}}  & \textbf{6.34 min}        & \textbf{7.16 s}                                                                    & \textbf{8.05 s}                                                               & \textbf{60.0 \%}          & \textbf{0.47}                                                         & \textbf{0.31}                                                          & \textbf{25.0 \%}     \\ \hline
\end{tabular}}
\caption{Performance summary of the experts who took the visual Turing test. A mean 60\% accuracy is close to the optimal 50\% accuracy that would indicate random guessing by the experts. On average the experts spent nearly the same amount of time studying the synthetic samples as they did with the real samples.}
\label{tab:table2}
\end{table*}

\begin{figure}[h!]
    \centering
    \includegraphics[width = 0.7\textwidth]{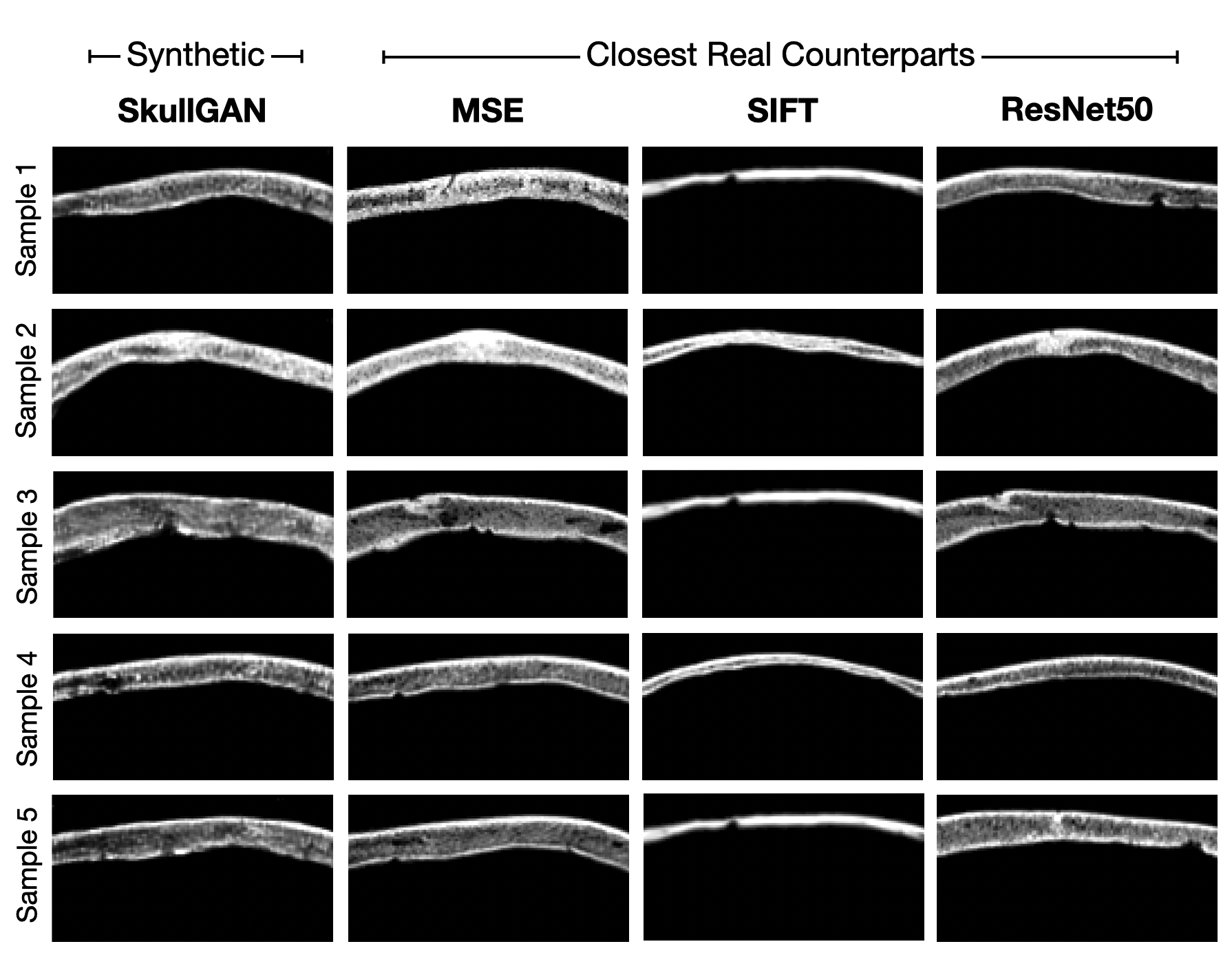}
    \caption{Example random SkullGAN samples and their closest real counterparts. Despite its simplicity, MSE, due to the fixed location of the skull segments, does reasonably well in identifying similar skull segments. SIFT seems to put too much weight on bright cortical bone and tends to select homogeneous, high HU-value segments. For samples 1, 3, and 5, SIFT has flagged the same closest real counterpart. Cosine similarity of the ResNet50 feature vectors, though computationally more intensive than MSE, shows a reliable performance in selecting similar segments.}
    \label{cross_sections}
\end{figure}

\subsection*{Memory GAN}
We employed three methods in identifying whether the generator in SkullGAN  was overfitting the training set: pixel-wise MSE, SIFT, and cosine similarity between the ResNet50 feature vectors. We found the candidates identified through pixel-wise MSE and ResNet50 to be visually more similar to one another, compared to candidates identified through SIFT. Examples are shown in Figure \ref{cross_sections}. None of the randomly generated samples were replicas of their closest real counterparts, allowing the conclusion that the SkullGAN network did not memorize the training set. 

\section*{Discussion}

In this work, we demonstrated the ability of SkullGAN to generate large numbers of synthetic skull CT segments that are visually and quantitatively quite similar to real skull CTs. The main advantage of using SkullGAN is its ability to overcome some of the challenges associated with obtaining real CT scans. Large datasets of anonymized, curated, and preprocessed medical images often are limited by factors such as time, capital, and access. In contrast, SkullGAN can generate an infinite number of highly varied skull CT segments quickly and at a very low cost. This makes it possible for any researcher to generate large datasets of skull CT segments for the purpose of training deep-learning models with applications involving the human skull, including but not limited to TUS. At the moment, when large samples of skull CTs are needed, either real human skull scans are manually preprocessed \cite{Miscouridou2022ClassicalSimulation}, or idealized models are simulated \cite{Stanziola2021AUltrasound, Hayner2001NumericalSkull}. However, manual preprocessing of real human skull CTs is viable when only a few samples are needed; idealized skulls, although easily generated in large numbers and capable of fooling the radiological metrics, are unlikely to yield high performance for supervised learning models. Instead of resorting to these idealized representations of skulls, researchers working on optimizing TUS simulation algorithms can now use SkullGAN to test their AI algorithms on large quantities of realistic synthetic skull CT segments.

One potential limitation of this work is that we have trained SkullGAN on a relatively small dataset of 38 healthy subjects. While the results are promising, it would be useful to test this technique on a larger and more diverse dataset to ensure that it generalizes well to other populations, accounting for variations in race, gender, and age. A larger dataset, on the order of hundreds of real human skull CTs, would also further improve the quality of the synthetic skull images.

Although SkullGAN only generates 2D slices of skull CT segments, most novel approaches in TUS are first verified and reported in 2D \cite{Stanziola2021AUltrasound, Wang2023Physics-informedPropagation}. This is inspired by the fact that if an algorithm fails in 2D it is bound to fail in 3D, and that 2D simulations are computationally less expensive. Therefore, although not yet ideal for clinical deep learning algorithms, SkullGAN is still immensely valuable for research and serves as a proof of concept for later 3D developments. Extending SkullGAN to generate 3D volumetric skull CT scans will require significantly more data points for training, beyond what we had available to us. 

In conclusion, our work presents a novel approach to address the main roadblock in integrating deep learning to the field of TUS, that is data scarcity, by generating large numbers of synthetic human skull CT segments. The results demonstrate that SkullGAN is capable of generating synthetic skull CT segments that are indistinguishable from real skull CT segments. Future work should investigate the performance of SkullGAN on larger and more diverse datasets, and extend SkullGAN to generate volumetric skull models. Much like ImageNet played a pivotal role in development of advanced deep learning algorithms in computer vision, by providing a very large labeled dataset for training, SkullGAN and its variants trained on other systems and organs of the human body \cite{Frid-Adar2018GAN-basedClassification, Baur2018GeneratingGANs, Skandarani2021GANsStudy, Khosravi2023CreatingConcerns} may play a similar role. Facilitated by models such as SkullGAN, the preponderance of such valid, high-quality, and preprocessed medical images readily available to any researcher will usher in a new wave of advanced deep learning models in healthcare, that go beyond classification and segmentation. 

\section*{Acknowledgements}

We would like to thank \conditionalCensor{Jeremy Irvin} and \conditionalCensor{Eric Luxenberg} for their helpful discussions, \conditionalCensor{Ningrui Li} and \conditionalCensor{Fanrui Fu} for advice on skull CT segmentation and preprocessing, and \conditionalCensor{Jeff Elias at The University of Virginia} for graciously providing 28 of the 38 human skull CTs for training SkullGAN. This work was generously supported by \conditionalCensor{NIH R01 Grant EB032743}.

\section*{Code Availability}

SkullGAN was written in Python v3.9.2 using PyTorch v1.9.0. All of the source code, training data, and the trained model are available at \conditionalCensor{https://github.com/kbp-lab/SkullGAN}.

\noindent\makebox[\linewidth]{\rule{\linewidth}{0.4pt}}

\bibliographystyle{ieeetr}

% \bibliography{references}
{\small \bibliography{references_new}}

\section*{Appendix}

\begin{figure*}[!htb]
    \centering
    \includegraphics[width = \textwidth]{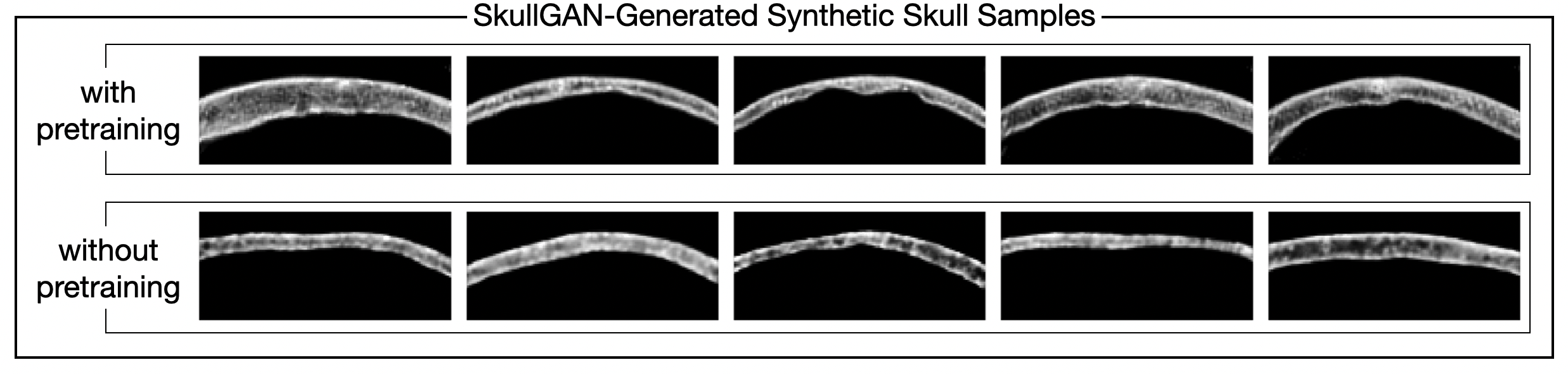}
    \caption{Effect of pretraining on the Celeb-A dataset. Pretraining on the Celeb-A dataset helped the model generate higher resolution skulls with finer edges and contours.} 
    \label{pretrain_effect}
\end{figure*}

\begin{figure*}[!htb]
    \centering
    \includegraphics[width = \textwidth]{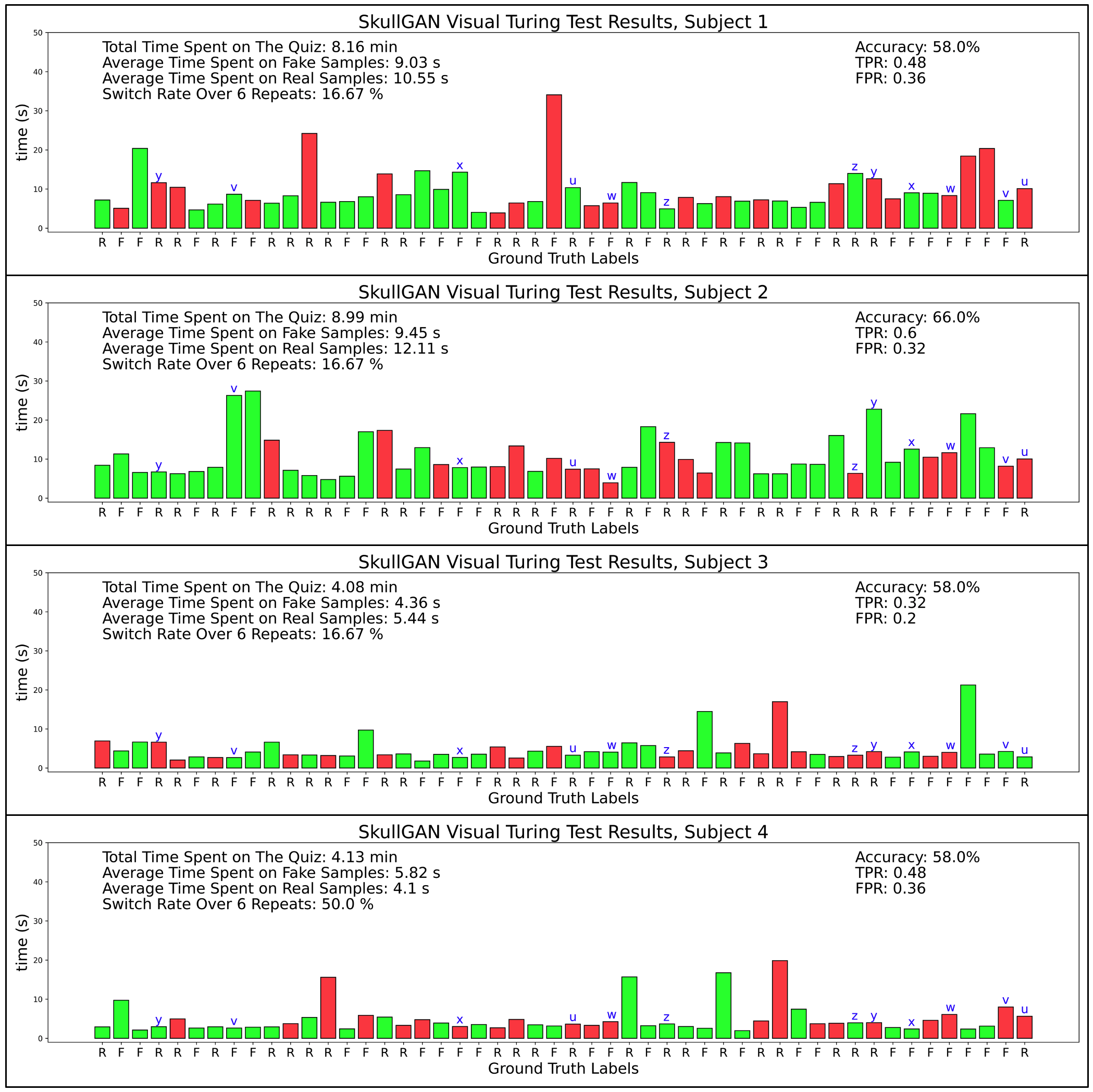}
    \caption{Visual Turing test performance. Detailed report of each subject's performance on the VTT is presented in each panel. The heights of the bars indicate the time spent on labeling each sample. Green color denotes correct labeling, while red color indicates incorrect labeling. Ground truth labels for each sample are shown along the x-axis. There are 3 pairs of duplicates in either category of real and synthetic/fake samples (a total of 12 pairwise-duplicate samples). The ratio of inconsistent labeling across these duplicate samples is reported as the switch rate. Letters \texttt{u, v, w, x, y}, and \texttt{z} indicate which samples were duplicates. }
    \label{SkullGAN-Quiz}
\end{figure*}

\end{document}